\documentclass[letterpaper,aps,prd,preprint,showpacs,nofootinbib,superscriptaddress]{revtex4}

\usepackage{epsf}
\usepackage{epsfig}
\usepackage{graphics}
\usepackage{graphicx}
\usepackage{amssymb}

\newcommand{\nc}{\newcommand}
\nc{\postscript}[2]{\setlength{\epsfxsize}{#2\hsize}\centerline{\epsfbox{#1}}}

\def\l{\left}
\def\r{\right}

\def\g{\gamma}

\def\ra{\rightarrow}
\def\d{\partial}

\nc{\beq}{\begin{equation}}   \nc{\eeq}{\end{equation}}
\nc{\bea}{\begin{eqnarray}}   \nc{\eea}{\end{eqnarray}}
\nc{\baa}{\begin{array}}      \nc{\eaa}{\end{array}}
\nc{\bit}{\begin{itemize}}    \nc{\eit}{\end{itemize}}
\nc{\ben}{\begin{enumerate}}  \nc{\een}{\end{enumerate}}
\nc{\bce}{\begin{center}}     \nc{\ece}{\end{center}}

\nc{\non}{\nonumber}

\begin{document}

\title{\begin{flushright}

\mbox{\normalsize \rm UMD-PP-10-010}\\
       \end{flushright}
\bf Higgs Production from Gluon Fusion in Warped Extra Dimensions}

\author{Aleksandr Azatov\footnote{aazatov@umd.edu}}
\affiliation{Maryland Center for Fundamental Physics, \\ Department of
  Physics, University of Maryland, \\ College Park, MD 20742, USA.
}

\author{Manuel Toharia\footnote{mtoharia@umd.edu}}
\affiliation{Maryland Center for Fundamental Physics, \\ Department of
  Physics, University of Maryland, \\ College Park, MD 20742, USA.
}

\author{Lijun Zhu\footnote{ljzhu@umd.edu}}
\affiliation{Maryland Center for Fundamental Physics, \\ Department of
  Physics, University of Maryland, \\ College Park, MD 20742, USA.
}

\begin{abstract}
We present an analysis of the loop-induced couplings of the Higgs
boson to the massless gauge fields (gluons and photons) in the warped
extra dimension models where all Standard Model fields propagate in the bulk. We show
that in such models corrections to the $hgg$ and $h\g\g$ couplings are
potentially very large. These corrections can lead to generically
sizable deviations in the production and decay rates of the Higgs
boson, even when the new physics states lie beyond the direct reach of the
LHC.

 \end{abstract}

\maketitle

\section{Introduction}

Warped extra dimensions, \textit{ \`{a} la} Randall-Sundrum model
(RS) present one of the most elegant solutions to the  Standard Model
(SM) hierarchy problem \cite{RS1}.
Placing SM fields in the bulk  of the extra dimension can
simultaneously explain the hierarchies of the SM fermion masses
\cite{a, Grossman, bulkSM}. Such models provide a very attractive way
to suppress flavor violation by the so called RS
Glashow-Iliopoulos-Maiani(GIM) mechanism \cite{a,
  AgashePerezSoni,hs}.
The electroweak precision tests put important bounds on the scale of new physics,
but by introducing custodial symmetries \cite{RSeff,Agashezbb} one can have it
around few TeV \cite{RSeff,EWPTmodel,Agashezbb}.

In this paper, we will analyze the Higgs couplings to massless vector
bosons in RS models where all SM fields are in the bulk, and
the modification to the $hgg$ and $h\g\g$ couplings arises from integrating
out Kaluza-Klein (KK) partners of the SM fields. Previous works on
this topic for RS models have been done in \cite{Lillie:2005pt,
  Djouadi:2007fm,Bouchart:2009vq,Casagrande:2010si,
  Cacciapaglia:2009ky}\footnote{One of the main differences between our
  work and previous analysis is that we present analytical results for
  the contribution of the full KK fermion tower. Other subtle differences
  are discussed in the main text.}. These effects were also studied
in models of warped extra dimensions in which the Higgs arises as
Pseudo-Nambu-Goldstone boson (PNGB) \cite{Falkowski:2007hz} and
within the effective theory formalism \cite{Giudice:2007fh,Low:2009di}. The studies of the Higgs production in flat extra dimensions in the models with gauge Higgs unification were carried out in \cite{Maru:2007xn}.  
We will stick to the models with flavor anarchy
\cite{AgashePerezSoni,hs} in which the hierarchies in masses and
mixings in the the fermion sector are explained by small overlap integrals between
fermion wave functions and the Higgs wave function along the extra
dimension. Previous studies of this framework have mainly focused on
bounds on the KK scale coming from new flavor violating sources.
In spite of the RS-GIM mechanism, it was still
found that $\Delta F=2$ processes  mediated by the KK gluon push the mass of the KK
excitations to be above $\sim 10$ TeV
\cite{Weiler,Fitzpatrick:2007sa,BlankedeltaF2},
making them very hard to produce and observe at the LHC
\cite{kkgluon}. These bounds coming from flavor
violation in low energy observables can be relaxed by introducing
additional flavor symmetries
\cite{Fitzpatrick:2007sa,Chen:2008qg,Cacciapaglia:2007fw,Csakifp},
or by promoting the Higgs to be a 5D bulk field (instead of being
brane localized) \cite{Agashe2site, Gedalia:2009ws}. A similar
tension was found in the lepton sector in \cite{Agashe:2006iy}, making
scale of $O(5)$ TeV still compatible with experiments. Lower KK scales
can be achieved by changing the fermion representations
\cite{Agashe:2009tu} or by introducing flavor symmetries
\cite{Chen:2008qg}. It is interesting to point out that flavor
violating effects can also be mediated by the radion \cite{Azatov:2008vm}, a graviscalar
degree of freedom which might be generically the lightest new physics
state and therefore may lead to important phenomenological bounds.
 More recently, it has also been pointed out that models
with fermions in the bulk give rise to flavor violation in the
couplings of Higgs to SM fermions
\cite{Agashe:2009di,Azatov:2009na}, 
leading to interesting constraints from $\Delta F=2$  processes
and to flavor violating collider signatures such as $h\ra t c$ (see also
the most recent analysis of \cite{Duling:2009pj, Casagrande:2010si} for further
details). Other interesting collider effects like rare top decays $t\ra cZ$ were discussed in \cite{Casagrande:2008hr}.

The outline of the paper is as follows:
in section II, we consider the
effect of just two vector-like heavy fermions, one
singlet under $SU(2)_L$ and one doublet. This simple case helps us
understand in simple terms the effects caused by the full
tower of KK fermions in a realistic 5D setup. In section III we
present a calculation of the $hgg$ and $h\g\g$
couplings for the simple model where all the fermions are in a doublet
representation of $SU(2)_L$ or $SU(2)_R$. In this section and in
Appendix \ref{appendixsumrules} we also present a simple way to
evaluate the complete KK fermion tower contribution to $hgg$ and
$h\g\g$ couplings.
Having explained and derived the new contributions to the Higgs couplings
caused by the heavy KK fermions, we proceed in
section IV to study quantitatively the main phenomenological effects
and outline our conclusions in section \ref{concl}.

\section{Warm-up: New Vector-Like Fermions}\label{warmup}

We  begin by computing the new contribution to the $hgg$ coupling using
effective theory with just the zero and first KK modes, where we only
consider one
family of light quarks (say, up and down quarks) augmented by the
presence of two heavy vector-like fermions, one in  doublet
representation of $SU(2)_L$ and the other in  singlet representation.
This effective theory description has the advantage of being
economical and gives lucid physical intuition of the source of new
physics contribution. Therefore, we adopt this approach in this
section just to illustrate the essential points of our
calculation. Moreover, the calculation is more general is the sense
that it applies to any Beyond Standard Model (BSM) model in which there
exist extra vector-like fermions which mix with SM fermions (see
\cite{Pierce:2006dh} for a similar discussion). The full calculation of
the $hgg$ coupling in the 5D warped extra dimension model will be carried out
in the next section.

\begin{figure}
	\includegraphics[scale=0.8]{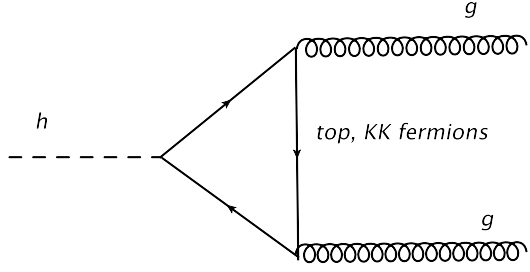}
\caption{$hgg$ coupling induced by fermion loop.\label{loop}}
\end{figure}

To start, we review here the Higgs boson production through gluon
fusion in SM. The coupling between gluon and Higgs mainly comes from
top quark loop (See Fig. \ref{loop}). The partonic cross section for
$gg \to h$ is \cite{higgshunter}
\begin{equation}
\sigma_{gg \to h}^{SM} = \frac{\alpha_s^2 m_h^2}{576 \pi }\left|\sum_Q \frac{y_Q}{m_Q} A_{1/2}(\tau_Q)\right|^2 \delta(\hat{s}-m_h^2),
\end{equation}
where the sum is for all SM fermions, $\hat{s}$ is invariant mass squared of the two incoming gluons, $\tau_Q \equiv m_h^2/4m_Q^2$, $y_Q$ and $m_Q$ are Yukawa couplings and masses of the quarks, and the form factor for fermion in the loop is
\begin{equation}
A_{1/2}(\tau) = \frac{3}{2}[\tau + (\tau-1)f(\tau)]\tau^{-2},
\end{equation}
where
\begin{equation}\label{Eq. ftau}
f(\tau) =[\arcsin\sqrt{\tau}]^2,\quad(\tau \le 1); \qquad -\frac{1}{4}\left[\ln\left(\frac{1+\sqrt{1-\tau^{-1}}}{1-\sqrt{1-\tau^{-1}}}\right) - i \pi\right]^2, \quad(\tau > 1).
\end{equation}

We note that for $\tau_Q \to 0$ i.e. $m_h \ll m_Q$, the form factor
tends to be unity, while for $\tau_Q \to \infty$ i.e. $m_h \gg m_Q$,
the form factor tends to zero. For reference, we consider a Higgs
boson with mass $120$ GeV, then for c-quark, we have $A_{1/2}(\tau_c)
\approx 0.01$; and for a KK fermion with mass $2000$ GeV, we have
$A_{1/2}(\tau_{kk}) \approx 1.00021 $. Therefore, it is a good
approximation to treat the form factors for KK fermions as unity,
while for light quarks, we can safely  ignore their contributions.

In the effective theory with just one KK mode, we have zero mode
fermions ($q_L,\, u_R$) and first KK fermions ($Q^{(1)}_L, Q^{(1)}_R,
U^{(1)}_L, U^{(1)}_R$), where $q, Q$ denote the up-type quark from
$SU(2)_L$ doublet, and $u, U$ denote the up-type quark from $SU(2)_L$
singlet.  Then we have the following mass matrix:
\begin{eqnarray}\label{massmat}
(\bar{q}_L, \bar{Q}^{(1)}_L, \bar{U}^{(1)}_L) \left(\begin{array}{ccc} \frac{Y_{q_L u_R} \tilde{v}}{\sqrt{2}}  & 0 & \frac{Y_{q_L U_R} \tilde{v}}{\sqrt{2}} \\ \frac{ Y_{Q_L u_R} \tilde{v}}{\sqrt{2}}& M_Q & \frac{Y_{Q_L U_R} \tilde{v}}{\sqrt{2}}  \\ 0 & \frac{Y_{U_L Q_R} \tilde{v}}{\sqrt{2}} & M_U \end{array}\right) \left( \begin{array}{c} u_R \\ Q^{(1)}_R \\ U^{(1)}_R \end{array} \right)+\hbox{h.c},
\end{eqnarray}
where $Y_{q_L u_R}$ etc. are the Yukawa couplings between the
corresponding chiral fermions, and $\tilde{v}$ is the Higgs VEV (note
that it is not the same as $v_{SM}$). The Yukawa couplings matrix is
given by
\begin{eqnarray}\label{yukawamat}
(\bar{q}_L, \bar{Q}^{(1)}_L, \bar{U}^{(1)}_L)  \left(\begin{array}{ccc} \frac{Y_{q_L u_R}}{\sqrt{2}} & 0 & \frac{Y_{q_L U_R}}{\sqrt{2}}  \\ \frac{Y_{Q_L u_R}}{\sqrt{2}} & 0& \frac{ Y_{Q_L U_R}}{\sqrt{2}}   \\ 0 &\frac{ Y_{U_L Q_R}}{\sqrt{2}}  & 0 \end{array}\right) \left( \begin{array}{c} u_R \\ Q^{(1)}_R \\ U^{(1)}_R \end{array} \right) h+\hbox{h.c}.
\end{eqnarray}

To calculate these fermion contributions to the $hgg$ coupling, we assume
that the masses of the KK fermions $\gg  m_h$, and therefore their form
factors are approximately unity.
Before proceeding let us classify different effects contributing to the shift of $hgg$ coupling from that of the SM:

\bit
\item relation between mass and Yukawa coupling of the lightest state (SM fermion) is modified from the SM value $y^{\text{light}}_{RS}\neq \frac{m_f }{v_{SM}}$;
\item we have loop of KK fermion running in the triangle diagrams (see Fig. \ref{loop}).
\eit
So we should calculate
\begin{eqnarray}\label{traceformula}
\frac{y^{\text{light}}_{RS}}{m^{\text{light}}} A_{1/2}(\tau_{\text{light}})+\sum_{\text{heavy}}\frac{Y_i}{M_i}=
\text{Tr}(\hat{Y}\hat{M}^{-1}) + \frac{y^{\text{light}}_{RS}}{m^{\text{light}}}\left(A_{1/2}(\tau_{\text{light}})-1\right),
\end{eqnarray}
where $\hat{M}$ and $\hat{Y}$ are the fermion mass and Yukawa matrices
given in Eq. (\ref{massmat}) and (\ref{yukawamat})\footnote{Note that the real part of the Yukawa coupling  will lead to the operator $h G_{\mu\nu}G^{\mu\nu}$, and the imaginary part will lead to the operator  $h G_{\mu\nu} \tilde G^{\mu\nu}$. For simplicity in this paper we everywhere will assume that we have only $h G_{\mu\nu}G^{\mu\nu}$ operator.}. The first term on
the LHS of the above equation gives the contribution from the SM
fermion (lightest mass eigenstate), and the second term comes from the
contributions of heavy KK fermions. Note that $\hat{Y} =
\frac{\partial \hat{M}}{\partial \tilde{v}}$, therefore, we can use
the following trick to calculate the trace \cite{Ellis:1975ap}:
\begin{equation}
\text{Tr}(\hat{Y} \hat{M}^{-1})= \text{Tr}\l(\frac{\d \hat{M}}{\d \tilde{v}} \hat{M}^{-1}\r) =\frac{\partial \ln \text{Det}(\hat{M})}{\partial \tilde{v}},
\end{equation}
we also have
\begin{equation}
\text{Det}(\hat{M}) = Y_{q_L u_R} M_Q M_U \frac{\tilde{v}}{\sqrt{2}} + Y_{Q_L u_R} Y_{U_L Q_R} Y_{q_L U_R}\l(\frac{\tilde{v}}{\sqrt{2}}\r)^3 - Y_{q_L u_R} Y_{Q_L U_R}Y_{U_L Q_R} \l(\frac{\tilde{v}}{\sqrt{2}}\r)^3 .
\end{equation}
Now  we expand to first order in $\frac{\tilde{v}^2}{M_Q M_U}$:
\begin{equation}
\text{Tr}(\hat{Y} \hat{M}^{-1}) \approx \frac{1}{\tilde{v}}\left[1 +  \left(\frac{Y_{Q_L u_R} Y_{U_L Q_R} Y_{q_L U_R}}{Y_{q_L u_R}} - Y_{Q_L U_R}Y_{U_L Q_R} \right) \frac{\tilde{v}^2}{M_Q M_U}\right].
\end{equation}
Note that the masses and Yukawa couplings of the SM fermions are also
modified (see \cite{Azatov:2009na} for details), 
\begin{eqnarray}
\frac{y^{\text{light}}_{RS}}{m^{\text{light}}} \approx \frac{1}{\tilde{v}} \left(1+ \frac{Y_{Q_L u_R} Y_{U_L Q_R} Y_{q_L U_R}}{Y_{q_L u_R}} \frac{\tilde{v}^2}{M_Q M_U} \right),
\end{eqnarray}
where the last expression was derived using the following assumption 
  ($Y_{q_L u_R}\ll Y_{q_LU_R},Y_{Q_Lu_R}\ll Y_{Q_LU_R}$). This assumption is true the quarks of the  first two generations, and the extra contribution which is important for the quarks of the third generation  will be presented in the next section.
  Now Eq. (\ref{traceformula}) reduces to
\begin{equation}\label{Eq. twosite result}
\frac{y^{\text{light}}_{RS}}{m^{\text{light}} } A_{1/2}(\tau_{\text{light}}) - \tilde{v} \frac{ Y_{Q_L U_R} Y_{U_L Q_R}}{M_Q M_U}.
\end{equation}
We can see that for the light generation quarks,
$A_{1/2}(\tau_{\text{light}}) \approx 0$, we get $ -
\frac{1}{\tilde{v}} Y_{Q_L U_R} Y_{U_L Q_R} \frac{\tilde{v}^2}{M_Q
  M_U}$, which is just the contribution coming from the KK modes. Note
that this contribution is proportional to $Y_{Q_L U_R}Y_{U_L
  Q_R}$, which is the product of Yukawa couplings of the KK fermions of opposite
chiralities, this structure of the contribution will become essential in calculating the effects in realistic warped model in the next section. It is interesting to see that even though the light SM
quarks give negligible contribution to $hgg$ coupling, their KK
partners can give sizable new contributions. In addition, there would
be an multiplicity enhancement of these KK contributions due to the
number of flavors.

The analysis above showed that additional vector-like fermions which
mix with SM fermions can alter the $hgg$ coupling significantly. In
warped extra dimension models with 5D fermions propagating in the
bulk, these extra vector-like fermions naturally come up as the KK
towers of fermions. Therefore, we expect generically sizable new
physics contributions to $hgg$ coupling in this class of models. We
carry out the detailed calculations in warped extra dimension in the
next section.

\section{Minimal Warped Extra Dimension Model with Custodial Protection}

In this section, we first calculate the KK fermion contributions to
$hgg$ coupling in warped extra dimensions (RS). We then apply similar
techniques to calculate both KK fermion and KK gauge boson
contributions to $h\g\g$ coupling. We show that simple analytical
formulas can be obtained for these new physics contributions.

\subsection{$hgg$ coupling in RS}
\label{3A}
In this subsection, we consider the effect of the full KK fermion
tower on $hgg$ coupling. We consider models with bulk gauge group
$SU(2)_L \otimes SU(2)_R$, which is motivated to ease the bound from
electroweak precision test \cite{RSeff}. We consider here just a
single family of quarks for the sake of simplicity. A generalization
to 3 generation quarks can be easily applied later. For the quark
fields, we consider the simple spinorial representation with the
following field contents:
\begin{eqnarray}
\left(\begin{array}{cc} Q_L^u (+, +)\, Q_R^u(-,-) \\ Q_L^d (+, +)\, Q_R^d(-,-) \end{array} \right), \quad  \left(\begin{array}{cc} U'_R (-, +)\, U'_L (+,-) \\ D_R (+, +)\, D_L (-,-) \end{array} \right),\quad \left(\begin{array}{cc} U_R (+, +)\, U_L (-,-) \\ D'_R (-, +)\, D'_L (+,-) \end{array} \right).
\end{eqnarray}
The first multiplet is a doublet of $SU(2)_L$ and the last two are
doublets of $SU(2)_R$. The boundary conditions are denoted for the
corresponding chirality. They have the following Yukawa
couplings \footnote{We consider here a general bulk Higgs
  \cite{bulkhiggs} with vector-like Yukawa coupling for simplicity.}
\begin{equation}
Y^u\sqrt{R} (\bar{Q}_L^u U_R + \bar{Q}_L^d D'_R)H + Y^d \sqrt{R} (\bar{Q}_L^u U'_R + \bar{Q}_L^d D_R)H + (L \leftrightarrow R) + \text{h.c. }
\end{equation}
Note that $Y^u$, $Y^d$ are dimensionless and order one, and $1/R=k$ is the curvature scale. After KK decomposition in the basis where Higgs vev is zero, we have zero modes $q_L^{u,{(0)}}, q_L^{d,{(0)}}, d_R^{(0)}, u_R^{(0)}$ and the KK modes $Q_{L,R}^{u,{(i)}}, Q_{L,R}^{d,{(i)}}, D_{L,R}^{(j)}, U_{L,R}^{(j)}, U^{\prime {(k)}}_{L,R}, D^{\prime {(k)}}_{L,R}$. For up-type quarks, we have the following infinite dimensional mass matrix
\begin{eqnarray}
(\bar{q}^{u,(0)}_L, \bar{Q}^{u,(i)}_L, \bar{U}^{(j)}_L, \bar{U}^{\prime (k)}_L) \left(\begin{array}{cccc} \frac{Y^u_{qu} \tilde{v}}{\sqrt{2}} & 0 &\frac{ Y^u_{qU_b} \tilde{v} }{{\sqrt{2}}} & \frac{Y^d_{qU'_c} \tilde{v}}{\sqrt{2}} \\ \frac{Y^u_{Q_iu} \tilde{v}}{\sqrt{2}} & M_Q & \frac{Y^u_{Q_iU_b} \tilde{v}}{\sqrt{2}} & \frac{Y^d_{Q_iU'_c} \tilde{v}}{\sqrt{2}} \\ 0 & \frac{Y^{u,*}_{U_jQ_a} \tilde{v}}{\sqrt{2}} & M_U & 0 \\ 0 & \frac{Y^{d,*}_{U'_kQ_a}\tilde{v}}{\sqrt{2}} & 0 & M_{U'} \end{array}\right) \left( \begin{array}{c} u_R^{(0)} \\ Q^{u,(a)}_R \\ U^{(b)}_R \\ U^{\prime (c)}_R \end{array} \right)+\hbox{h.c},
\end{eqnarray}
where $i,j,k,a,b,c$ are KK indices. The Yukawa couplings matrices are defined e.g. by
\bea
Y_{Q_iU_b}^u = Y^u\sqrt{R}\int dz\l(\frac{R}{z}\r)^5 h(z) q_L^{u,(i)}(z) u_R^{(b)}(z),
\eea
 i.e. it is an integral of product of Higgs and fermion wavefunctions, where $h(z)$ is a profile of the Higgs field normalized in  the following way
\bea
1=\int_R^{R'}dz \l(\frac{R}{z}\r)^3 h(z)^2.
\eea
 The KK mass matrices are diagonal, e.g. $M_Q = \text{diag} (M_{Q_1},
 M_{Q_2}, \cdots)$.
One naively might think that the couplings $Y_{U_jQ_a}$ vanish in the
limit of brane Higgs due to the odd boundary conditions of $U_L$ and
$Q^u_R$, so it is safe to ignore them in this matrix. But these are
precisely the $Z_2$ odd operators described in detail in
\cite{Azatov:2009na} (detailed analysis without these operators was
presented in \cite{Bouchart:2009vq}). These operators as was shown in
\cite{Azatov:2009na} 
    lead to flavor violation in the
     Higgs sector, and they are also essential in evaluating the $hgg$
     coupling\footnote{These operators can be mimicked by higher
       dimensional derivative operators \cite{Azatov:2009na}, which
       shows UV sensitivity of the effect.}. To avoid subtleties with
     wave function being discontinuous at IR brane we will assume that
     the Higgs is 5D bulk field and only at the end we will take a
     brane Higgs limit.

 Now we can use the same determinant trick, the determinant of the mass matrix to the order of $\tilde{v}^3$ is
\begin{eqnarray}
&&\text{Det}(\hat{M}) =  \left(\prod_{i,j,k} M_{Q_i} M_{U_j} M_{U'_k} \right) \times\nonumber\\
&&\l[ \frac{Y^u_{qu} \tilde{v}}{\sqrt{2}} - Y^u_{qu} \l(\frac{\tilde{v}}{\sqrt{2}}\r)^3\sum_{a,b}\l(
\frac{ Y^d_{Q_a U'_b} Y^{d,*}_{U'_b Q_a}}{M_{Q_a} M_{U'_b}}
+ \frac{Y^u_{Q_a U_b} Y^{u,*}_{U_b Q_a} }{M_{Q_a} M_{U_b}} \r)\r.
\nonumber\\
 &&+\l.
\l(\frac{\tilde{v}}{{\sqrt{2}}}\r)^3 \sum_{a,b} \l( \frac{Y^u_{q U_b}Y^{u,*}_{U_b Q_a} Y^u_{Q_au}}{M_{Q_a} M_{U_b}} +  \frac{Y^d_{q U'_b}Y^{d,*}_{U'_b Q_a}  Y^u_{Q_a u} }{M_{Q_a} M_{U'_b}}\r)\r].
\end{eqnarray}
Now we get
\bea 
\text{Tr}(\hat{Y}\hat{M}^{-1}) = \frac{\partial \ln \text{Det}(\hat{M})}{\partial \tilde{v}}=\frac{1}{\tilde{v}}\l[1-\tilde{v}^2\sum_{a,b}\l(
\frac{ Y^d_{Q_a U'_b} Y^{d,*}_{U'_b Q_a}}{M_{Q_a} M_{U'_b}}
+ \frac{Y^u_{Q_a U_b} Y^{u,*}_{U_b Q_a} }{M_{Q_a} M_{U_b}} \r)
\r.\nonumber\\
\l.
+\frac{\tilde{v}^2}{Y_{qu}^u} \sum_{a,b} \l( \frac{Y^u_{q U_b}Y^{u,*}_{U_b Q_a} Y^u_{Q_au}}{M_{Q_a} M_{U_b}} +  \frac{Y^d_{q U'_b}Y^{d,*}_{U'_b Q_a}  Y^u_{Q_a u} }{M_{Q_a} M_{U'_b}}\r)\r].
\eea
Again, for the light generation quarks there are corrections to the SM fermion masses and Yukawa couplings \cite{Azatov:2009na}
\begin{eqnarray}
\label{mandyuk}
m^{\text{light}} &=& Y^u_{qu} \frac{\tilde{v}}{\sqrt{2}} + \sum_{a,b}Y^u_{qU_b}\frac{1}{M_{U_b}} Y^{u,*}_{U_b Q_a} \frac{1}{M_{Q_a}}Y^u_{Q_a u} \l(\frac{\tilde{v}}{\sqrt{2}}\r)^3 \\ \nonumber &+& \sum_{a,b} Y^d_{qU'_b}\frac{1}{M_{U'_b}} Y^{d,*}_{U'_b Q_a} \frac{1}{M_{Q_a}} Y^u_{Q_a u} \l(\frac{\tilde{v}}{\sqrt{2}}\r)^3 ,\nonumber\\
y^{\text{light}}_{RS} &=& \frac{Y^u_{qu}}{{\sqrt{2}}} + \frac{3}{{\sqrt{2}}}\sum_{a,b}Y^u_{qU_b}\frac{1}{M_{U_b}} Y^{u,*}_{U_b Q_a} \frac{1}{M_{Q_a}}Y^u_{Q_a u} \l(\frac{\tilde{v}}{\sqrt{2}}\r)^2 \\ \nonumber &+& \frac{3}{{\sqrt{2}}}\sum_{a,b}Y^d_{qU'_b}\frac{1}{M_{U'_b}} Y^{d,*}_{U'_b Q_a} \frac{1}{M_{Q_a}} Y^u_{Q_a u} \l(\frac{\tilde{v}}{\sqrt{2}}\r)^2 .
\end{eqnarray}
Therefore
\begin{equation}\label{ytopovermtop}
\frac{y^{\text{light}}_{RS}}{m^{\text{light}}} \approx \frac{1}{\tilde{v}}\left(1  + \sum_{a,b}Y^u_{Q_au}Y^{u,*}_{U_b Q_a} Y^u_{q U_b} \tilde{v}^2 \frac{1}{M_{Q_a} M_{U_b} Y^u_{qu}}+ \sum_{a,b} Y^u_{Q_a u} Y^{d,*}_{U'_b Q_a} Y^d_{q U'_b} \tilde{v}^2 \frac{1}{M_{Q_a} M_{U'_b} Y^u_{qu}}  \right) .
\end{equation}
So the total contribution to $hgg$ coupling by light generation quarks and their KK partners  is (see Eq. \ref{traceformula})
\begin{equation}
\label{hggcoup}
-\tilde{v}\sum_{a,b}\l(
\frac{ Y^d_{Q_a U'_b} Y^{d,*}_{U'_b Q_a}}{M_{Q_a} M_{U'_b}}
+ \frac{Y^u_{Q_a U_b} Y^{u,*}_{U_b Q_a} }{M_{Q_a} M_{U_b}} \r)+ \frac{y^{\text{light}}_{RS}}{m^{\text{light}} } A_{1/2}(\tau_{\text{light}}).
\end{equation}
Note that this result is very similar to the one we obtained in the last section (Eq. \ref{Eq. twosite result}), except for an extra term corresponding to the contribution of extra states in the doublet representation of $SU(2)_R$. For light generations, the  last term is negligible, and we are left with first two terms. The first two terms can be written as
\begin{eqnarray}\label{convolutionintegral}
-\tilde{v}\sum_{a,b}\Big[ Y^u Y^{u,*} R \left(\int {dz}{dz'} \l(\frac{R}{z}\r)^5 \l(\frac{R}{z'}\r)^5 \frac{q_L^{(a)}(z)q_R^{(a)}(z')}{M_{Q_a}} \frac{u_R^{(b)}(z)u_L^{(b)}(z')}{M_{U_b}} h(z) h(z') \right) \\ \nonumber
+ Y^d Y^{d,*} R  \left( \int dzdz' \l(\frac{R}{z}\r)^5 \l(\frac{R}{z'}\r)^5 \frac{q_L^{(a)}(z)q_R^{(a)}(z')}{M_{Q_a}} \frac{u_R^{\prime (b)}(z)u_L^{\prime (b)}(z')}{M_{U'_b}} h(z)h(z') \right)\Big].
\end{eqnarray}
Now we have to evaluate the following sums
\bea \label{Eq. kk sum}
\sum_{a>0}\frac{q_L^{(a)}(z)q_R^{(a)}(z')}{M_{Q_a}},\quad \sum_{b>0} \frac{u_R^{(b)}(z)u_L^{(b)}(z')}{M_{U_b}}, \quad \sum_{b>0} \frac{u_R^{\prime (b)}(z)u_L^{\prime (b)}(z')}{M_{U'_b}}.
\eea
We can calculate them by using equations of motion for fermion
wavefunctions (see discussion in the  Appendix
\ref{appendixsumrules}). From the forms of these sums (see
Eq. (\ref{sumrules})), we see that we need to evaluate the integrals
of Higgs wavefunction times $\theta(z-z')$ and $\theta(z-z')^2$. This
can be done for general bulk Higgs. But for illustration purpose we
take the brane Higgs limit of bulk Higgs. Then we get
\begin{eqnarray}
\int dz dz' \theta(z-z')^2 h^{brane}(z)h^{brane}(z') = \frac{1}{2},
\end{eqnarray}\footnote{ To evaluate this integral we have to somehow
  regularize the wavefunction of the brane Higgs($\delta$ function),
  we used bulk Higgs inspired regularization of the delta function
  $h^{brane}(z)=\displaystyle \lim_{\beta\ra\infty}
  \frac{\beta}{R'}\l(\frac{z}{R'}\r)^\beta$.
 One can also use a rectangular regularization of brane Higgs wavefunction which will lead to the same result.}
and Eq. (\ref{convolutionintegral}) now reduces to
\begin{eqnarray}
 \frac{1}{2}\l(Y^uY^{u,*}+Y^dY^{d,*}\r) \tilde{v} R^{\prime 2}.
\end{eqnarray}
Therefore, for light generations, the contribution to $hgg$ coupling
is  $\l(Y^uY^{u,*}+Y^dY^{d,*}\r) \tilde{v} R'^2/2$, which comes just from KK
fermions and is independent of fermion bulk mass parameters. 

For the third generation quarks there will be an extra contribution to the formula in (Eq. \ref{ytopovermtop}) which we parameterize following \cite{Azatov:2009na} as $(-\frac{\Delta^{t,b}_2}{m \tilde{v}})$ (see Appendix \ref{appendixC} for details). This gives us additional contribution relative to (Eq. \ref{hggcoup}) 
\begin{equation}
\frac{\Delta^t_2}{m_t\tilde {v}}+\frac{\Delta^b_2 }{m_b\tilde{v}}.
\end{equation}
Also in this case contributions of the SM bottom and top qaurks are no longer negligible, so we have to include them
\bea
\frac{y_{b}^{RS}}{m_{b}} A_{1/2}(\tau_{b})+\frac{y_{t}^{RS}}{m_{t}} A_{1/2}(\tau_{t}).
\eea
Note that now  Yukawa couplings of the top and bottom quarks are
shifted( see discussion in  Appendix {\ref{appendixC}}).

It is simple to generalize the above result to three generations. The
KK towers of the quarks give a contribution proportional to
$\text{Tr}(Y_uY_u^\dagger + Y_dY_d^\dagger )$, and we have to combine
them with the effect coming from top and bottom quarks. To summarize,
compared with SM, the Higgs production cross-section from gluon fusion
in RS is
\begin{eqnarray}\label{Eq. hgg ratio}
\frac{\sigma^{RS}_{gg \to h}}{\sigma^{SM}_{gg\to h}} = \left(\frac{v_{SM}}{\tilde{v}}\right)^2\left|\frac{\text{Tr}(Y_u Y_u^\dagger + Y_d Y_d^\dagger){\tilde{v}^2 R'^2} +\frac{\Delta^t_2}{m_t}+\frac{\Delta^b_2}{m_b}+ x_t A_{1/2}(\tau_{t}) +x_b A_{1/2}(\tau_{b})}{A_{1/2}(\tau_{t})+A_{1/2}(\tau_{b})}\right|^2 ,
\end{eqnarray}
where $x_t = \frac{y_t^{RS}\tilde{v}}{m_t}$ and $x_b =
\frac{y_b^{RS}\tilde{v}}{m_b}$ , with $y_t^{RS}$,$y_b^{RS}$
the shifted top and bottom Yukawa couplings in RS (reference
\cite{Casagrande:2010si} presented numerical results for the analysis
of the brane Higgs model including $Z_2$ odd operators, however, it is
hard to compare it with our result due to different particle content
of the models). We consider here the ratio $\frac{\sigma^{RS}_{gg \to
    h}}{\sigma^{SM}_{gg\to h}}$ in order to reduce the uncertainty
coming from higher order QCD corrections.
It is also important to notice that in the case when the couplings of
the $SU(2)_L$ and $SU(2)_R$ are not equal the ratio
$\left(\frac{v_{SM}}{\tilde{v}}\right)$ might be quite significant,
see discussion and analysis in \cite{Bouchart:2009vq}. In the rest of
the paper we will assume that $SU(2)_L$ and custodial $SU(2)_R$  have
the same gauge couplings (see appendix \ref{appendixB} for discussion
of VEV shift in this case).

It is also interesting to point out that the same diagrams that
contribute to the gluon fusion will also contribute to the
modification of the di-Higgs production. This might become an
interesting option to disentangle new physics contribution (see
discussion in the effective field theory approach in
\cite{Pierce:2006dh}).

\subsection{$h\g\g$ coupling in RS}
The calculation  of the $h\g\g$ coupling comes from  similar diagrams
as the one for the $hgg$ coupling, the only difference now is that we
have to take into account contributions of the towers of charged KK
gauge bosons and KK leptons. We will again use the simplest custodial
model where  leptons are in the doublet representation of  $SU(2)_L$
or $SU(2)_R$. We can calculate their contribution in the same way as
we did for the quarks.
Contribution of the KK tower of the $W_\pm$ was presented in
\cite{Bouchart:2009vq}, so here we just quote their results and the
reader can find more details about the derivation in the Appendix
\ref{appendixB}. The contribution of the tower of the KK $W_\pm$ is
given by
 \bea
\sum_{n \geq 0} \frac{C^n_{diag}}{2M^2_n} A_1(\tau_n)
 =\frac{C_{hww} }{2M^2_w} (A_1(\tau_w)+7)-\frac{7}{\tilde v},
  \eea
 where $C^n_{diag}$ is coupling between Higgs field and the n-th KK
 modes (mass eigenstates) of the $W_{\pm}$, and $C_{hww}$ is coupling between SM $W$ and
 the Higgs. $A_1(\tau_w)$ is the form-factor for the gauge bosons (see
 Eq. (\ref{Eq. A1form})).
 Including the modification of the coupling between SM $W$ and Higgs,
 this sum can be expresssed in the following way:
  \bea
  \sum_{n \geq 0} \frac{C^n_{diag}}{2M^2_n} A_1(\tau_n)&&=\frac{g^2 \tilde v}{4M_w^2}\l(1-\frac{\tilde v^2 R'^2(g_{5D}^2+\tilde g_{5D}^2)}{4 R}\r)(A_1(\tau_w)+7)-\frac{7}{\tilde v}\nonumber\\
&&\approx\frac{1}{\tilde  v}\l[
\l(1-\frac{\tilde v^2 R'^2(g_{5D}^2+\tilde g_{5D}^2)}{8 R}\r)A_1(\tau_w)
-\frac{7}{8}\tilde v^2 R'^2\frac{(g_{5D}^2+\tilde g_{5D}^2)}{R}\r],
 \eea
where $g_{5D}$ and $\tilde g_{5D}$ are the 5D gauge couplings of
$SU(2)_L$ and $SU(2)_R$ respectively. Adding both fermion and gauge
boson contributions together, now we can present our results for the
ratio of $\Gamma(h \to \g\g)$ between RS and SM:
\bea
\label{hgammagamma}
&&\frac{\Gamma^{RS}(h\ra \g\g)}{\Gamma^{SM}(h\ra \g\g)}=\l(\frac{v_{SM}}{\tilde v}\r)^2 \frac{1}
{|A_1(\tau_w)+\frac{16}{9}A_{1/2}(\tau_t) + \frac{4}{9} A_{1/2}(\tau_b)|^2}
 \nonumber\\
&&\l|\l(1-\frac{ v_{SM}^2 R'^2(g_{5D}^2+\tilde g_{5D}^2)}{8R}\r)A_1(\tau_w)-\frac{7 v_{SM}^2 R'^2(g_{5D}^2+\tilde g_{5D}^2)}{8R}+\frac{16}{9}x_t A_{1/2}(\tau_t)\r.\nonumber\\
&&\l. +\frac{4}{9}x_b A_{1/2}(\tau_b)
+ \frac{1}{2}v^2_{SM} R'^2\text{Tr}\l[\frac{20}{9}\l(Y_u^\dagger Y_u +Y_d^\dagger Y_d\r)+\frac{4}{3}Y_l^\dagger Y_l\r]
+\frac{16\Delta^t_2}{ 9m_t}+\frac{4 \Delta^b_2}{9 m_b}\r|^2.
\eea

\section{Phenomenology}

In this section, we discuss the phenomenology of the Higgs boson in warped
extra dimensions. We focus our study on the Higgs production through
gluon fusion and the branching fraction of $h \to \g\g$ decay. We will
compare our results with that of holographic PNGB Higgs model studied
in \cite{Falkowski:2007hz}.

\begin{figure}
	\includegraphics[scale=1.2]{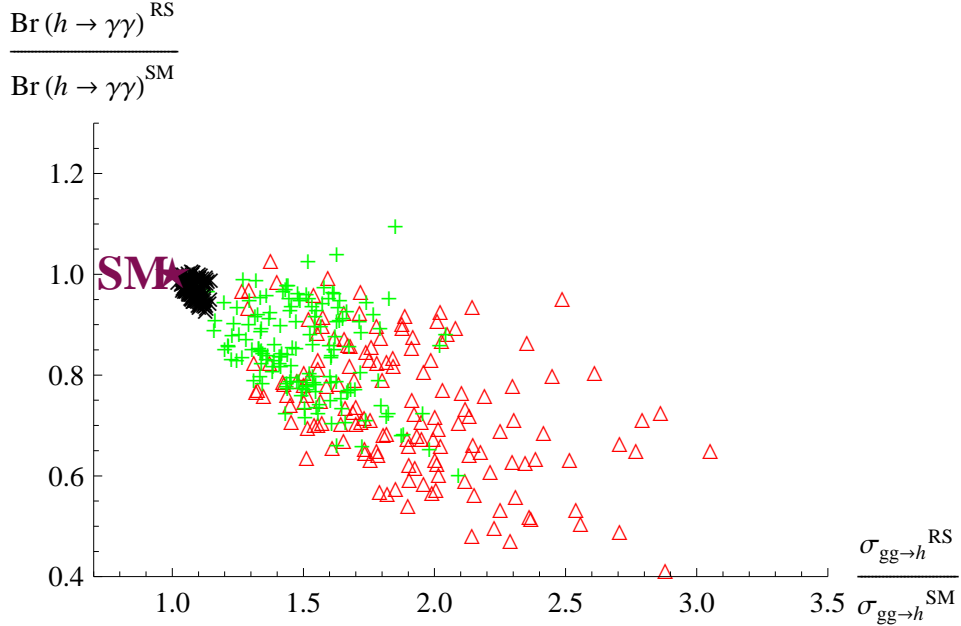}
\caption{\label{plotbulk}
 Scattered plot of  $\frac{\sigma_{gg\ra h}^{RS}}{\sigma_{gg\ra
     h}^{SM}}$ and
 $\frac{\text{Br}(h\ra\g\g)^{RS}}{\text{Br}(h\ra\g\g)^{SM}}$, for bulk
 Higgs with vector-like Yukawa couplings ($Y_1=Y_2$). The
 dimensionless 5D Yukawa couplings are varied between $Y\in[0.3,3]$
 and $m_h=120$ GeV. The black ``$\times$'' corresponds to the KK scale
 $R'^{-1}=5$ TeV, green ``$+$'' to $R'^{-1}=2$ TeV, and red
 ``$\triangle$'' to $R'^{-1}=1.5 $ TeV. The SM value is marked by the
 star.}
\end{figure}

To get a handle on the size of new physics contributions, we scan the
parameter space of RS with the assumption of flavor anarchy, i.e. the
5D Yukawa matrices are order one and uncorrelated. We find the set of
5D Yukawa couplings and fermion zero mode wavefunctions which give the
correct SM quark masses and CKM mixing. We then calculate
$\sigma(gg\to h)$ and $\text{Br}(h \to \g\g)$ using Eq. (\ref{Eq. hgg
  ratio}) and (\ref{hgammagamma}), and find the ratio
with that of SM. The result of the scan for bulk Higgs is shown in
Fig. \ref{plotbulk}.

We can see from the plot in Fig. \ref{plotbulk} that the new physics
contribution to $\sigma(gg\to h)$ tends to be positive and gets larger
for lower KK scale. Also the new physics contribution to $\sigma(gg\to
h)$ and $\text{Br}(h \to \g\g)$ are correlated: an increase in
$\sigma(gg\to h)$ is accompanied by a decrease in $\text{Br}(h \to
\g\g)$.
Before proceeding further let us stop and see whether we can
understand these results intuitively. First let us focus on the
enhancement of the Higgs production due to gluon fusion.
As we argued in the sections \ref{3A} these effects come mainly from
the modification of the top Yukawa coupling and from the loop with KK
fermions. As was shown in \cite{Azatov:2009na} top Yukawa coupling is
reduced compared to the SM value, so naively one should expect the
reduction of the Higgs production. But let us now look on the
contribution of the KK modes. One can see from (Eq.  \ref{Eq. hgg
  ratio}) that this contribution is proportional to  $\text{Tr}(Y_u
Y_u^\dagger+Y_d Y_d^\dagger)$ which is always positive, so the sign
of this contribution is fixed. Also the typical size of  this term
will be roughly equal to $N^2 \bar Y^2$ where N is number of SM
families and $\bar Y$ is an average size of the Yukawa couplings, so
adding both up and down quark KK towers will lead to an overall
enhancement factor of 18.\footnote{One can see that  for  sufficiently
  large Yukawa couplings our expansion in powers of
  $YY^\dagger v^2 R'^2$ might become ill defined, and also
  contribution of the  higher order loops with KK fermions and Higgs
  might become important, so the one loop result becomes not reliable
  if the new physics contribution is much larger than that of the
  SM. At the same time we would like to note that our result even for
  the large 5D Yukawa couplings will give a typical size of the
  expected correction to the SM coupling.
} Therefore KK fermions give a large positive contribution to
$\sigma(gg\to h)$. Reduction of the $\text{Br}(h\to\g\g)$ can be
understood from the fact that in the SM the dominant contribution
comes from the loop with $W^\pm$, and the fermion contribution has an
opposite sign, thus enhancement of the fermion contributions
effectively decreases the overall coupling.

This implication is two-fold. First, it means that even with a KK scale
out of the reach of the LHC ($\gtrsim 5$ TeV), we can still probe the
framework of warped extra dimension by precision measurements of
various Higgs production and decay processes. Second, by comparing our
result with that of \cite{Falkowski:2007hz}, we can see that
$\sigma(gg\to h)$ can be used to distinguish between RS with bulk
Higgs and holographic PNGB Higgs model (or gauge-Higgs
unification). In the latter model, a reduction is usually expected,
which can be contrasted with our results for bulk Higgs. Note that the
difference in these two models comes from the extra symmetry in PNGB
Higgs, which constrains the Higgs interactions (see discussion in
\cite{Giudice:2007fh, Low:2009di}).

\begin{figure}[htb]
	\includegraphics[scale=1.2]{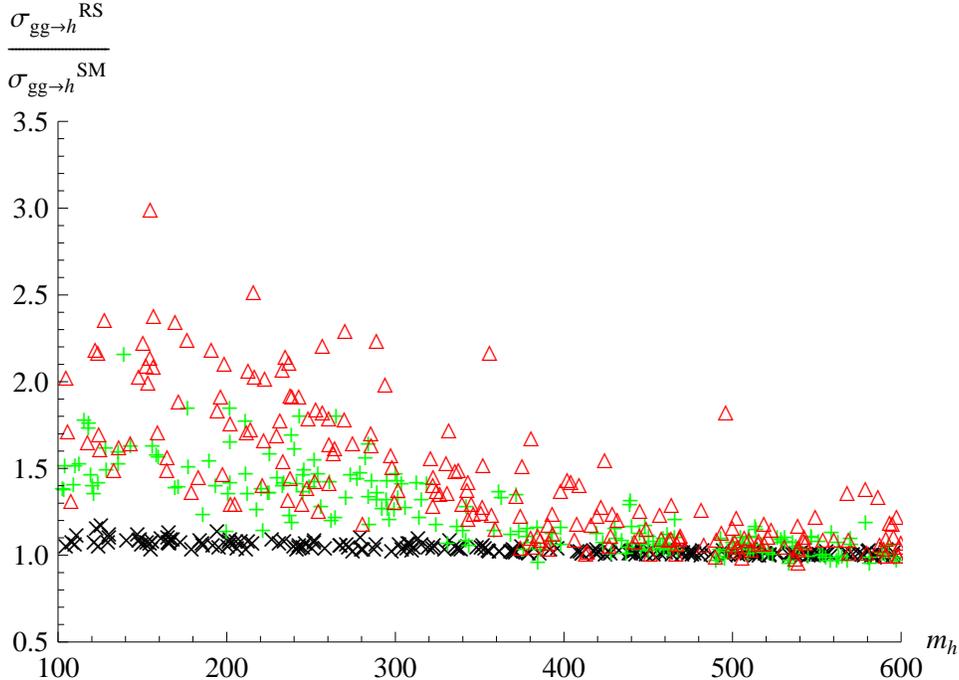}
\caption{\label{Higgs mass dependence} Dependence of
  $\frac{\sigma_{gg\ra h}^{RS}}{\sigma_{gg\ra h}^{SM}}$ on the Higgs
  mass  for different values of  $R'^{-1}$ in bulk Higgs scenario with
  vector-like Yukawa couplings ($Y_1=Y_2$). The dimensionless 5D
  Yukawa couplings are varied between $Y\in[0.3,3]$.  The black
  ``$\times$'' corresponds to KK scale $R'^{-1}=5$ TeV, green ``$+$''
  to  $R'^{-1}=2$ TeV, and  red ``$\triangle$'' to $R'^{-1}=1.5$
  TeV. }
\end{figure}

\begin{figure}[htb]
	\includegraphics[scale=1.2]{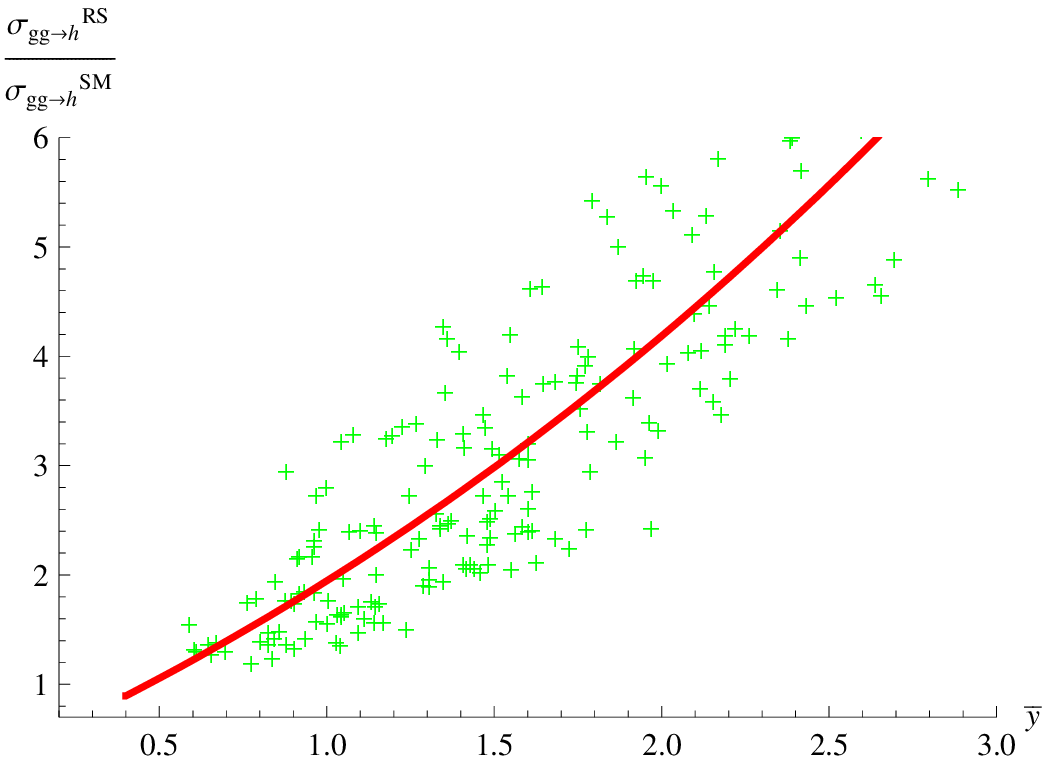}
\caption{\label{plotydependence} Dependence of   $\frac{\sigma_{gg\ra
      h}^{RS}}{\sigma_{gg\ra h}^{SM}}$ on the average size of
  dimensionless 5D Yukawa couplings $\bar Y$, for the Higgs mass
  $m_h=120$ GeV and KK scale $R'^{-1}=2$ TeV.}
\end{figure}

To study the dependence of new physics contributions on the Higgs
boson mass, we plot in Fig. \ref{Higgs mass dependence} the ratio
$\frac{\sigma_{gg\ra h}^{RS}}{\sigma_{gg\ra h}^{SM}}$ vs. $m_h$ for
various KK scales. We can see that the new physics contribution
decreases as $m_h$ increase from $100$ to $\sim 360$ GeV. This can be
understood from the fact that in SM, the form factor for the top quark
attains its largest value when $m_h \approx 2 m_t$. Since in RS with
bulk Higgs, the top quark Yukawa coupling is reduced compared to that
of SM, there is a larger negative new physics contribution to $hgg$
coupling when $m_h \approx 2 m_t$, leading to a smaller total new
physics contribution.

In Fig. \ref{plotydependence}, we plot the dependence of the ratio
$\frac{\sigma_{gg\ra h}^{RS}}{\sigma_{gg\ra h}^{SM}}$ on the average
size of the 5D Yukawa couplings. We can see quite clearly that the
size of new physics contribution increases as the 5D Yukawa couplings
increases. This is expected from the fact that KK fermion
contributions are proportional to $\text{Tr}(Y_u Y_u^\dagger+Y_d
Y_d^\dagger)$. In the framework of flavor anarchy, the 5D Yukawa
couplings are order one. We can see from Fig. \ref{plotydependence}
that for order one Yukawa couplings, we have sizable new physics
contributions to $\sigma(gg\ra h)$.

\begin{figure}[htb]
	\includegraphics[scale=1.2]{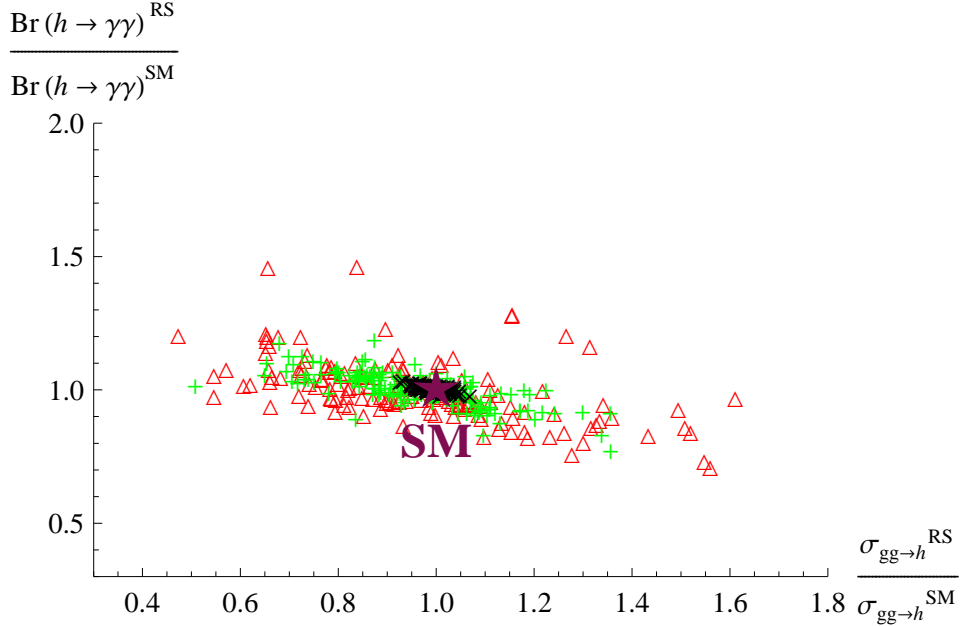}
\caption{\label{plotbrane}Scattered plot for the modification of
  $\frac{\text{Br}(h\ra\g\g)^{RS}}{\text{Br}(h\ra\g\g)^{RS}}$ and
  $\frac{\sigma_{gg\ra h}^{RS}}{\sigma_{gg\ra h}^{SM}}$ for brane
  Higgs with $Y_1$ independent of $Y_2$, where 5D Yukawa couplings are
  varied between $Y\in[0.3,3]$ and $m_h=120$ GeV. The black
  ``$\times$'' corresponds to the KK scale $R'^{-1}=5$ TeV, green
  ``$+$'' to the  $R'^{-1}=2$ TeV, and  red ``$\triangle$'' to the
  $R'^{-1}=1.5$ TeV. The SM value is marked by the star.}
\end{figure}

So far we have been assuming that the Higgs is the bulk field and 5D Yukawa couplings are vector-like i.e.
\bea
{\cal L}= Y_1 \bar Q^u_L U_R H +Y_2 \bar U_L Q^u_R H \quad\text{with } Y_1=Y_2.
\eea
In the case where the Higgs is a 5D bulk field this condition of $Y_1=Y_2$
is forced by the 5D Lorentz symmetry. But the Higgs can be brane localized
or even a bulk Higgs might have brane localized couplings and these
couplings do not have to respect 5D bulk Lorentz symmetry. So
generally speaking $Y_1\neq Y_2$, and they could be independent of
each other. Let us see how this might modify our results. The first
thing to notice is that the contribution of the tower of  KK modes now
has the following structure $Y_1 Y_2^{\dagger}$. Before proceeding
further we immediately see that the overall sign of the contribution
is not fixed any more! So we cannot predict in generic RS model the
sign of the effect: whether it is enhancement or suppression for both
$hgg$ and $h\g\g$ couplings. This is shown in Fig. \ref{plotbrane}. We
can see that the size of new physics contribution is generically large
for moderate KK scale, but now its sign can be both positive and
negative.

\section{Conclusions}
\label{concl}
In conclusion, we summarize the results presented in the paper. We
calculated the corrections to the  $hgg$ and $h\g\g$ couplings in RS at
one loop order. We have found that the new physics states can modify
significantly these couplings. We have shown that the dominant
contribution to these coupling comes from the towers of KK fermions
running inside triangle diagrams. We have shown that the KK towers of
the light fermions do contribute significantly to these couplings,
contrary to the models with Higgs being a PNGB boson where  this
contribution is sub-leading. We have shown that in the
 models with the Higgs in the bulk  and Yukawa couplings being
 vectorlike ($Y_1 = Y_2$), $hgg$ coupling becomes enhanced and $h\g\g$
 coupling  suppressed compared to that of SM, even though the top
 Yukawa coupling is suppressed compared to the SM value. This naively
 counterintuitive result is explained by the fact that
 the contribution of the KK towers of all SM fermions  is so strong
 that it overcomes the effect from suppression of the top Yukawa
 coupling.
 Modification of the Higgs production cross-section remains
 significant even for a KK scale far from LHC
 accessibility. Specifically,  we can get order one  corrections even
 with lightest KK modes above 5 TeV. For the generic
models with Higgs on the brane or bulk Higgs with brane Yukawa
couplings the sign of the effect remains unpredictable. We might have
enhancement as well as suppression, but the parametric size of the effect
remains the same. The total effect comes from collective contributions
of the KK partners of all generations. Therefore, the size of these
new physics contributions is large, even if the KK fermions are
heavy. This shows us that in the absence of new resonances  an
analysis of the Higgs couplings might become a very important tool  in
understanding the structure of BSM physics.

\section*{Acknowledgements}

We would like to thank Kaustubh Agashe for his encouragement, comments
and suggestions. We would also like to thank Matthias Neubert for
useful discussion and Uli Haisch, Florian Goertz, Sandro Casagrande and Torsten Pfoh for comments. A.A. and L.Z. would like to thank Maryland Center
for Fundamental Physics for support during the completion of this
project.

\appendix
\section{KK sum rules }\label{appendixsumrules}
In this section we will present a way of efficiently performing KK sums for the fermions \footnote{Similar tricks were discussed in \cite{Hirn:2007bb}} (such as Eq. (\ref{Eq. kk sum})). Let us look at the equations of motions for the fermions in the absence of the Higgs vev. In the absence of the Higgs vev we can always choose a basis where 5D bulk masses are diagonal, and so we can ignore all the mixings. Let us concentrate on the KK decomposition
of the $SU(2)_L$ doublet $Q_{L,R}$ with boundary
conditions $(\pm, \pm)$. The equations of motion of the KK wavefunctions are
\begin{eqnarray}
-m_{n} q^{(n)}_L - \partial_z q_R^{(n)} + \frac{c_q+2}{z}q_R^{(n)}  = 0,\\
-m_{n}^* q_R^{(n)} + \partial_z q_L^{(n)} + \frac{c_q-2}{z}q_L^{(n)}  = 0.
\end{eqnarray}
We  take the first equation and rewrite it as:
\bea
-m_{n} q_L^{(n)} - z^{c_q+2}\d_z\left(q_R^{(n)}\ z^{-c_q-2} \right)=0.
\eea
We now multiply by $z^{-c_q-2}$ and integrate between $R$ and $z_1$:
\bea
\label{eqm1}
-m_{n} \int_R^{z_1}dz  z^{-c_q-2} q_L^{(n)}(z)  &=&
q_R^{(n)}\ z^{-c_q-2}\left.\vphantom{\int^A_A}\right|_R^{z_1} ,  \nonumber\\
  \int_R^{z_1}dz  z^{-c-2} q_L^{(n)}(z)  &=&-\frac{1}{m_{n}}
q_R^{(n)}(z_1)\ z_1^{-c-2}.
\eea
We now use the completeness relation
\bea
&&\sum^\infty_{n=0} q_L^{(n)}(z_2)q_L^{(n)}(z) =\frac{z^4}{R^4} \delta(z_2-z)\\
\Rightarrow&& \sum^\infty_{n=1} q_L^{(n)}(z_2)q_L^{(n)}(z) =\frac{z^4}{R^4} \delta(z_2-z)- q_L^0(z_2)q_L^0(z).
\eea
Based on (Eq \ref{eqm1})  we will get
\bea
-\int_R^{z_1}dz\ z^{-c_q-2} \sum^\infty_{n=1} q_L^{(n)}(z_2)q_L^{(n)}(z) = z_1^{-c_q-2} \sum_{n=1}^\infty  \frac{q_R^{(n)}(z_1)q_L^{(n)}(z_2)}{m_{n}},
\eea
where we have explicitly extracted the zero mode contribution from the
sum. Let us note that
\bea
q_L^0(z)=N_L z^{2-c_q} \ \ \ {\rm with} \ \ \ \ N_L=\sqrt{\frac{1-2c_q}{\epsilon^{2c_q-1}-1}} R^{c_q-5/2},
\eea
and where we have defined the warp factor $\epsilon =
\frac{R}{R^\prime} \sim 10^{-16}$.

 Now we can finally write:
\bea
\sum_{n=1}^\infty  \frac{q_R^{(n)}(z_1)q_L^{(n)}(z_2)}{m_{n}}&=&
-z_1^{c+2}\int_R^{z_1}dz\ z^{-c-2}\Big( \frac{z^4}{R^4} \delta(z_2-z)-
q_L^0(z_2)q_L^0(z) \Big)\nonumber\\
&=&
\frac{z_1^{2+c_q}z_2^{2-c_q}}{R^4}\left[-\theta(z_1-z_2)+ \frac{\l (\frac{z_1}{R}\r)^{1-2c}-1}{\l(\frac{R'}{R}\r)^{1-2c}-1} \right].
\eea
Similarly we can calculate the sum for the other three possible boundary conditions :
\bea
\label{sumrules}
&&\psi_L(+,+):~~\sum\frac{q_R^{(n)}(z_1)q_L^{(n)}(z_2)}{m_{n}}=
\frac{z_1^{2+c}z_2^{2-c}}{R^4}\l[ -\theta(z_1-z_2)+\frac{\l(\frac{z_1}{R}\r)^{1-2c}-1}{\l(\frac{R'}{R}\r)^{1-2c}-1}\r ],\nonumber\\
&&\psi_L(+,-):~~
\sum\frac{q_R^{(n)}(z_1)q_L^{(n)}(z_2)}{m_{n}}=-\frac{
z_1^{2+c}z_2^{2-c}}{R^4}\theta(z_1-z_2),
\nonumber\\
&&\psi_L(-,+):~~
\sum\frac{q_R^{(n)}(z_1)q_L^{(n)}(z_2)}{m_{n}}=
\frac{z_1^{2+c}z_2^{2-c}}{R^4}\theta(z_2-z_1),\nonumber\\
&&\psi_L(-,-):~~
\sum\frac{q_R^{(n)}(z_1)q_L^{(n)}(z_2)}{m_{n}}=
\frac{z_1^{2+c}z_2^{2-c}}{R^4}\l[\theta(z_2-z_1)-\frac{\l(\frac{z_2}{R}\r)^{1+2c}-1}{\l(\frac{R'}{R}\r)^{1+2c}-1}\r].
\eea
Using these relations we can now perform all the necessary sums to calculate the KK fermion contribution to $hgg$ coupling.

\section{Gauge boson couplings and contribution to $h\g\g$ coupling}
\label{appendixB}
In this section just for the sake of the completion we  present analysis for the modification of the gauge boson coupling to the Higgs boson, and their contribution to the $h\g\g$ coupling. We start from the modification of the Higgs vev
\bea
v_{SM}^2\approx\tilde v^2-\frac{\tilde v^4 R'^2}{8 R}\l(g_{5D}^2+\tilde g_{5D}^2\r),
\eea
where $v_{SM}=246$ GeV, $\tilde g_{5D}$ is five dimensional gauge coupling of the custodial $SU(2)_R$, so
\bea
\tilde v\approx v_{SM} \l( 1+\frac{R'^2 v_{SM}^2}{16 R} (g_{5D}^2+\tilde g_{5D}^2)\r).
\eea
This effect will lead to the overall modification of the SM $hgg$ and $h\g\g$ coupling by the factor
$1-\frac{R'^2 v_{SM}^2}{16 R} (g_{5D}^2+\tilde g_{5D}^2)\approx 0.95$
for ($R'^{-1}=1500\, \hbox{TeV}, g_{5D}=\tilde g_{5D}
$).

\subsection{Couplings of $W^\pm$ to Higgs in RS}
To calculate modification of the $h\g\g$ coupling we also have to calculate contribution coming from  the $W$ boson. From the Lagrangian (see \cite{RSeff})
\bea
{\cal  L}=\frac{g^2}{2}
\l(\l(\frac{h+\tilde v}{\sqrt{2}}\r)^2-\frac{R'^2(g_{5D}^2+\tilde g_{5D}^2)}{4R}\l(\frac{h+\tilde v}{\sqrt{2}}\r)^4\r)W^+_\mu W^{-\mu},
\eea
one can immediately deduce coupling between Higgs and $W$.
\bea
&&{\cal L} =C_{hww}\, h W^+_\mu W^{-\mu},\nonumber\\
&&C_{hww}=\frac{g^2 \tilde v}{2}\l[1-\frac{R'^2 (g_{5D}^2+\tilde g_{5D}^2) \tilde  v^2}{ 4 R}\r].
\eea

\subsection{Contribution of the KK tower of $W_\pm$ to the $h\g\g$}
In this subsection we  derive the contribution of the $W_\pm$ KK modes to the $h\g\g$ coupling
(we will closely follow discussion presented in \cite{Bouchart:2009vq}). First let us denote by $M^2$ the mass squared matrix of the charged gauge bosons, then the coupling to the Higgs boson will be given by the matrix $$ C=\frac{\d M^2}{\d \tilde{v}},$$ rotating back to the basis where mass matrix $M^2$ is diagonal we will get
\bea
C_{diag}=U\frac{\d M^2}{\d \tilde{v}}U^\dagger,
\eea
where $U$ is a unitary matrix that diagonalizes $M$.
We can parameterize the contribution of the gauge boson KK modes to the $h\g\g$ coupling in the following way:
 \bea
 \sum_{n \geq 0} \frac{C^n_{diag}}{2M^2_n} A_1(\tau_n)=\frac{C_{hww}}{2M^2_w} A_1(\tau_w)+\sum_{n>0}\frac{C^{n}_{diag}}{2M^2_n} A_1(\tau_n),
 \eea
where $A_1(\tau)$ is the form factor for vector bosons in the loop  \cite{higgshunter} ($\tau=m_h^2/4M_n^2$)
\begin{equation}\label{Eq. A1form}
A_1(\tau) = - [2\tau^2 + 3\tau + 3(2\tau-1)f(\tau)]\tau^{-2},
\end{equation}
where $f(\tau)$ is given by Eq. (\ref{Eq. ftau}). For KK gauge bosons $\tau_n \to 0$, and  $A_1(\tau_n) \approx -7$,  so we get
 \bea \label{Eq. kkw gamma}
\frac{C_{hww} }{2M^2_w} A_1(\tau_w)-7\sum_{n>0}\frac{C^{n}_{diag} }{2M^2_n} &=&\frac{C_{hww}}{2M^2_w} (A_1(\tau_w)+7)-7\sum_{n\geq 0}\frac{C^{n}_{diag}}{2M^2_n}\,.
 \eea
To evaluate $\displaystyle\sum_{n\geq0}\frac{C^{n}_{diag}}{2M^2_n}$ we can use the following trick \cite{Ellis:1975ap}
\bea
\sum_{n \geq 0} \frac{C_{diag}^n}{M_{n}^2}=\text{Tr}\l[\l(M_{diag}^2\r)^{-1} C \r]=\text{Tr}\l[\frac{\d M^2}{\d \tilde{v}}\l(M^2\r)^{-1} \r]=\frac{\d}{\d \tilde{v}} \ln \l(\text{Det} M^2\r).
\eea
Let us see how the determinant of the gauge boson mass matrix depends on $\tilde v$. For simplicity  we assume that
the Higgs is localized on the IR brane. We denote by $f_{(i)},\tilde f_{(j)}$ values of the profiles on the IR brane for KK modes of $SU(2)_L$ and $SU(2)_R$ gauge bosons respectively. Then the mass matrix will look like:
\bea
M^2=\l(\baa{cccc}g_{5D}^2 f_{(0)}^2\frac{\tilde v^2}{4}&f_{(0)}f_{(1)} g_{5D}^2 \frac{\tilde v^2}{4}& f_{(0)} \tilde f_{(1)} g_{5D} \tilde{g}_{5D}\frac{\tilde v^4}{2} &...\\
g_{5D}^2 f_{(0)} f_{(1)} \frac{\tilde v^2}{4}&M_1^2+ {f_{(1)}}^2 g_{5D}^2 \frac{\tilde v^2}{4}& f_{(1)} \tilde f_{(1)} g_{5D} \tilde g_{5D}\frac{\tilde v^2}{4} &...\\
g_{5D} \tilde g_{5D} f_{(0)} \tilde f_{(1)} \frac{\tilde v^2}{4}& f_{(1)} \tilde f_{(1)} g_{5D} \tilde g_{5D} \frac{\tilde v^2}{4}&\tilde M_1^2 + {{\tilde{f}}_{(1)}}^2 {\tilde{g}_{5D}}^2\frac{\tilde v^2}{4} &...\\
\vdots&\vdots&\vdots&\ddots\\
\eaa\r).
\eea
One can see from the structure of the matrix that the determinant is equal to
\bea
\text{Det} M^2=g_{5D}^2 f_{(0)}^2\frac{\tilde v^2}{4} \prod_{i,j} M_i^2 \tilde M_j^2.
\eea
We have checked that for generic bulk Higgs $\text{Det} M^2\propto \tilde{v}^2+ O(\tilde{v}^6) $, one can calculate it using mixed position momentum propagators. So the results presented in this section are approximately independent of the Higgs localization. Now we can proceed to the evaluation of the sum in Eq. (\ref{Eq. kkw gamma}) and substituting result for the determinant we get
\bea
\sum_{n \geq 0} \frac{C^n_{diag}}{2M^2_n} A_1(\tau_n)
 &=&\frac{C_{hww} }{2M^2_w} (A_1(\tau_w)+7)-\frac{7}{\tilde v}.
\eea

\section{Review of Higgs Flavor violation}
\label{appendixC}
In this appendix we  present general formulas for the misalignment between SM fermion masses and  Higgs Yukawa couplings in RS(see for details\cite{Azatov:2009na}). We define the following quantity to parameterize the misalignment
\bea
\hat{\Delta}=\hat{m}-{\tilde{v}} \hat{y},
\eea
where $\hat{m},\hat{y}$ are mass matrix and Yukawa couplings of the SM fermions. Then it can be split into two parts
\bea
\hat{\Delta}=\hat{\Delta}_1+\hat{\Delta}_2,
\eea
where $\hat{\Delta}_1$ is the main contribution for the light generations and $\hat{\Delta}_2$ becomes important only for the third generation of quarks.
Then calculations show that $\hat{\Delta}_1$ for the up type quarks is equal to
\bea
\hat{\Delta}^u_1=\frac{\tilde v\sqrt{2}}{3} \l(\frac{\tilde v^2 R'^2}{2}\r)\hat{F}(c_q)\l[Y_uY_u^{\dagger}Y_u+Y_dY_d^{\dagger}Y_u\r]\hat{F}(-c_u)
\eea\footnote{We assume here that Yukawa couplings are vectorlike $Y_2=Y_1$}
where $c_u,c_q$ are bulk mass parameters for the multiplets containing zero modes of the SM right-handed and left-handed up quarks respectively. $\hat{F}(c)$ is a diagonal matrix with elements given by the profiles of the corresponding quarks respectively 
\bea
F(c)\equiv\sqrt{\frac{1-2c}{1-\l(\frac{R}{R'}\r)^{1-2c}}}.
\eea 
One can get these expressions by evaluating the sum (Eq. \ref{mandyuk}) directly using the rules of (Eq. \ref{sumrules}) or by solving for the exact wavefunctions profiles as described in \cite{Azatov:2009na}. For the other contribution $\hat{\Delta}_2$ we will get
the following expression
\bea
\hat{\Delta}^u_2=R'^2\l[\hat{m}_u
\l( \hat{m}_u^{\dagger}
\hat{K}(c_q)+\hat{K}(-c_u)\hat{m}_u^{\dagger}
\r)m_u+\hat{m}_d\hat{\tilde{K}}(-c_d)\hat{m}_d^{\dagger}\hat{m}_u\r]
\eea
where
\bea
\tilde{K}(c)&\equiv&\frac{1-\l(\frac{R'}{R}\r)^{2c-1}}{1-2c}\frac{1-\l(\frac{R'}{R}\r)^{-2c-1}}{1+2c},\nonumber\\
K(c)&\equiv&
\frac{1}{1-2c}\l[-\frac{1}{\l(\frac{R}{R'}\r)^{2c-1}-1}+\frac{\l(\frac{R}{R'}\r)^{2c-1}-\l(\frac{R}{R'}\r)^2}{\l(\l(\frac{R}{R'}\r)^{2c-1}-1\r)(3-2c)}+\frac{\l(\frac{R}{R'}\r)^{1-2c}-\l(\frac{R}{R'}\r)^2}{(1+2c)\l(\l(\frac{R}{R'}\r)^{2c-1}-1\r)}\r]\nonumber\\
\eea
Note that subdominant contribution $\Delta_2$ is only important for the third generation, and in the text we denote $\Delta^{t,b}_2$ to be equal to $(\hat{\Delta}^{u,d}_2)_{33}$.
\end{document}